\documentclass[showpacs,preprintnumbers,amssymb,nofootinbib,aps]{revtex4}
\usepackage{epsfig,color,wrapfig}
\usepackage{amsmath}
\usepackage{graphicx}
\usepackage{amssymb}
\usepackage{slashbox}
\usepackage{float}
\usepackage{bm}% bold math
\usepackage{dcolumn,multirow}% Align table columns on decimal point
\usepackage[font={small}]{caption, subfig}

\newcommand{\ba}{\begin{array}}
\newcommand{\ea}{\end{array}}
\def\beq{\begin{equation}}
\def\eeq{\end{equation}}
\def\bea{\begin{eqnarray}}
\def\eea{\end{eqnarray}}
\def\nn{\nonumber}

\def\roughly#1{\mathrel{\raise.3ex\hbox
{$#1$\kern-.75em\lower1ex\hbox{$\sim$}}}}
\def\lsim{\roughly<}
\def\gsim{\roughly>}
\def\sla#1{\raise.15ex\hbox{$/$}\kern-.57em #1}% Feynman slash

\def\bd{B_d^0}

\def\order{\lower 1.8ex \hbox{\LARGE\~{}}}

\def\bdtau{B\to D^{(\ast)}\tau\nu_{\tau}}
\def\rdast{{{\cal BR}{(B\to D^{(\ast)}\tau\nu_{\tau})}}/{{\cal BR}{(B\to D^{(\ast)}\ell\nu_{\ell})}}}

\def\bdell{B\to D^{(\ast)}\ell\nu_{\ell}}
\def\bd0tau{B\to D \tau\nu_{\tau}}
\def\bdasttau{B\to D^{\ast}\tau\nu_{\tau}}
\def\bdastell{B\to D^{\ast}\ell\nu}
\def\be {\begin{equation}}
\def\ee {\end{equation}}
\def\diff{d{\cal B}/{dq^2}}

\begin{document}
%\preprint{UdeM-GPP-TH-11-202 }
\title{Optimal-observable analysis of possible new physics in $B\to D^{(\ast)}\tau\nu_{\tau}$ }

\author{Srimoy Bhattacharya}
\affiliation{Indian Institute of Technology, North Guwahati, Guwahati 781039, Assam, India }

\author{Soumitra Nandi}
\affiliation{Indian Institute of Technology, North Guwahati, Guwahati 781039, Assam, India }

\author{Sunando Kumar Patra}
\affiliation{Indian Institute of Technology, North Guwahati, Guwahati 781039, Assam, India }

 \begin{abstract}
 We study all possible observables in $\bdtau$ with new physics (NP), including new vector, scalar and
 tensor interactions, and investigate the prospects of extracting NP Wilson
 coefficients with optimal observables. Analysis of the full $q^2$ integrated branching fractions of $\bdtau$ show that the
 overall sensitivity of the observables of $\bd0tau$ is more towards the scalar current, whereas the bin-by-bin analysis 
 of $q^2$ distribution of the differential branching fraction points to regions of
 $q^2$ sensitive to tensor interactions. Interestingly, the observables in
 $\bdasttau$ are more sensitive to tensor interactions, and bin-by-bin analysis of this mode shows the distinct
 regions of $q^2$ sensitive to vector or scalar interactions. In addition to that, the $\tau$ polarisation asymmetry is 
 found to be more sensitive to NP compared to the other observables, in both decay modes.
 \end{abstract}

\maketitle

\section{Introduction}
Measurements of branching fractions and other related observables in semileptonic decays of 
$B$ meson to $\tau$ can be interesting for an indirect probe of NP. Earlier measurements on 
$R(D^{(\ast)})=\rdast$ by Belle \cite{Adachi:2009qg} and BABAR collaborations \cite{Lees:2013uzd} have shown 
some deviations from their Standard Model(SM) predictions \cite{Fajfer:2012vx}, indicating a possible signature of 
NP in $b\to c \tau \nu_{\tau}$ transitions. 
Several authors have tried to explain the observation in various NP scenarios \cite{Tanaka:1994,Sakaki:2013,Duraisamy:2014},
as well as in a model-independent way  \cite{Tanaka:2012,Sakaki:2014,Freytsis:2015}.  
In order to distinguish between the possible signatures of NP, the study of NP in $q^2$
distribution of differential branching fractions in $\bdtau$, various correlations among 
$\tau$ forward-backward asymmetry and $\tau$ polarization asymmetry has been examined 
in the Ref. \cite{Fajfer:2012,Sakaki:2013,Biancofiore:2013}.
Recently, a 2.1$\sigma$ deviation in the measurement of $R(D^{\ast})$ has been reported by 
LHCb collaboration \cite{Aaij:2015yra}; Belle collaboration has also announced
their most recent results on $R(D^{(\ast)})$  and measured values are consistent with the SM within error 
bars \cite{Huschle:2015}. 

So far, the constraints on the new couplings are obtained assuming their presence one at a time 
\cite{Sakaki:2013,Sakaki:2014}. If we consider all the interactions together, then it will be an impossible
task to extract all the couplings from a single measurement. However, if one can reduce the number of coupling parameters
by imposing certain constraints on the full set of parameters, only then it is possible to obtain meaningful
errors on the couplings, although the information lost due to various assumptions cannot be
retrieved. Therefore, it will be useful to have independent couplings, parametrised in such a way that the
measured errors on different parameters are uncorrelated. On the other hand, it is not necessary for a particular
observable to have equal sensitivity to different types of NP operators. Therefore, it is useful to know how an observable
can be optimised to guarantee
the maximal sensitivity to a particular type of NP interaction, which in turn will help us select observables
suitable for the extraction of a particular type of coupling. Hence, from a phenomenological
point of view, it is important to find out the significance of  
different types of NP interaction, to an observable. To achieve this goal, we use the optimal-observable analysis using
the invariant mass squared $q^2$, of the lepton-neutrino system. We construct
the optimal observable to identify the NP structure that can be best extracted from a particular observable, with
reasonable statistics. It also provides a deeper understanding of the sensitivity that can be best
obtained by any method, for a certain process. This technique has been widely used in
collider phenomenology \cite{davier93,Diehl:1993br,Atwood:1991ka,Gunion:1996,Diehl:1997,wudka2004}.

In this article, we analyse the $q^2$ distribution of the differential branching fractions, 
$R(D^{(\ast)})$, the $\tau$ polarisation asymmetries, forward backward asymmetries of the decay
$\bdtau$, and $D^*$ polarisation asymmetry. We include all possible non-standard four-fermi effective interactions of 
the lowest dimension, and estimate the expected statistical uncertainties in the extraction 
of various NP Wilson coefficients that can contribute to $\bdtau$.

\section{Methodology}

The most general effective Hamiltonian describing the $b\to c\tau \nu_{\tau}$ transitions with 
all possible four-fermi operators in the lowest dimension is given by \cite{Sakaki:2014}
\begin{align}
\nn {\cal H}_{eff} &= \frac{4 G_F}{\sqrt{2}} V_{cb}\Big[( 1 + C_{V_1}) {\cal O}_{V_1} +
 C_{V_2} {\cal O}_{V_2} \\
 &+ C_{S_1} {\cal O}_{S_1} + C_{S_2} {\cal O}_{S_2}+ C_{T}{\cal O}_{T}\Big],
 \label{eq1}
\end{align}
where the operator basis is defined as
\bea
{\cal O}_{V_1} &=& ({\bar c}_L \gamma^\mu b_L)({\bar \tau}_L \gamma_\mu \nu_{\tau L}) \nn,~ 
{\cal O}_{V_2} = ({\bar c}_R \gamma^\mu b_R)({\bar \tau}_L \gamma_\mu \nu_{\tau L}) \nn, \\
{\cal O}_{S_1} &=& ({\bar c}_L  b_R)({\bar \tau}_R \nu_{\tau L}) \nn, ~~~~~~~~
{\cal O}_{S_2} = ({\bar c}_R b_L)({\bar \tau}_R \nu_{\tau L}) \nn, \\
{\cal O}_{T}   &=& ({\bar c}_R \sigma^{\mu\nu} b_L)({\bar \tau}_R \sigma_{\mu\nu} \nu_{\tau L}),
\label{eq2}
\eea
and the corresponding Wilson coefficients are given by $C_W (W=V_1,V_2,S_1,S_2,T)$.
In this basis, neutrinos are assumed to be left handed. Our main focus is on 
the $q^2$ distribution of differential decay rate $d\Gamma/dq^2$ in $\bdtau$. The complete expressions 
are given in ref.\cite{Sakaki:2013}.

As mentioned earlier, the optimal-observable analysis is a technique to systematically estimate the statistical
uncertainties of the measurable parameters while extracting them from some observable. Elaborate discussions
on this technique can be found in references \cite{Diehl:1993br,Atwood:1991ka,Gunion:1996,Diehl:1997}.
In order to apply this technique to $\bdtau$, it is necessary to express the $q^2$ distribution of the differential
decay rate as
\be
\frac{d{\Gamma}(\bdtau)}{dq^2} =  \sum\limits_{i} C_i f_i(q^2),
\label{eq3}
\ee
where $C_i$s are functions of $C_W$s. The theoretical expressions for $C_i$s, along with the $f_i(q^2)$s, can be 
extracted from a direct comparison between the similar terms on both sides of eq.(\ref{eq3}). 
The coefficients $C_i$, relevant for the branching fractions in $\bdtau$, are given in Table \ref{cird}, 
and the corresponding $f_i(q^2)$s are given in the Appendix (Table \ref{fisrd}). 

\begin{table*}[!htbp]
\begin{center}
\begin{tabular}{|c|c|c|}
\hline
\cline{2-3}
\backslashbox{$C_i$}{Obs} & $\diff$ in $\bd0tau$ & $\diff$ in $\bdasttau$    \\
\hline
$C_1$ & $|1 + C_{V_1} + C_{V_2}|^2$ & $|1 + C_{V_1}|^2 + |C_{V_2}|^2$    \\
\hline
$C_2$ & $| C_{S_1} + C_{S_2}|^2$ & $Re[(1 + C_{V_1})C_{V_2}^{\ast}]$   \\
\hline
$C_3$ & $|C_{T}|^2$  &  $|C_{S_1} - C_{S_2}|^2$   \\
\hline
$C_4$ & $Re[(1 + C_{V_1} + C_{V_2})(C_{S_1}^{\ast} + C_{S_2}^{\ast})]$ & $|C_{T}|^2$   \\
\hline
$C_5$ & $Re[(1 + C_{V_1} + C_{V_2})C_T^{\ast}]$ & $Re[(1 + C_{V_1} - C_{V_2})(C_{S_1}^{\ast} - C_{S_2}^{\ast})]$  \\
\hline
$C_6$ &   --    & $Re[(1 + C_{V_1})C_T^{\ast}]$    \\
\hline
$C_7$ &   --    & $Re[C_{V_2} C_T^{\ast}]$      \\
\hline
\end{tabular}
\end{center}
%\label{cird}
\caption{$C_i$s as defined in eq.(\ref{eq3}). The observable $P^R_{\tau} (q^2)$ contains the same set of $C_i$s.}
\label{cird}
\end{table*}
The goal of this technique is to extract $C_i$s, which can be done by defining 
suitable weighting functions $w_i(q^2)$ such as $C_i = \int w_i (q^2) ({d\Gamma}/{dq^2}) dq^2$. In general various 
choices of $w_i$s are possible. However, there is a unique choice for which the resulting error in the extraction of  
$C_i$ is minimized \footnote{$C_i$s are minimised in a sense that the whole 
covariance matrix is at a stationary point in terms of varying the 
functional forms of $w_i(q^2)$ while maintaining $\int{w_i(q^2) f_k(q^2)} = \delta_{ik}$.}, and these functions are given by 
\be
w_i(q^2) = \sum_j \frac{X_{ij} f_j(q^2)}{d{\Gamma}/{dq^2}},
\label{wi}
\ee
where $X_{ij}$ is the inverse of $M_{ij}$ which is defined as 
%using the functions $f_i(q^2)$:
%
\be
 M_{ij} = \int dq^2 \frac{f_i(q^2) f_j(q^2)}{f_{SM}(q^2)}.
\label{eq4}
\ee
In the above expression, $f_{SM}(q^2)$ can be obtained from eq.(\ref{eq3}) by setting $C_W = 0$, while $C_{SM} = 1$. 
Hence, using eqs. (\ref{wi}), and (\ref{eq4}) the statistical uncertainties in $C_i$ extracted from the 
branching fractions can be obtained as \cite{Atwood:1991ka,Gunion:1996}
\be
|\delta C_i| = \sqrt{\frac{X_{ii} {{\cal B}(\bdtau)}^{exp}}{N_{sig}}} = \sqrt{\frac{X_{ii}}{\sigma_P {\cal L}_{eff}}},
\label{eq5}
\ee
where ${\cal B}^{exp} = (1/\Gamma) \int dq^2  d\Gamma/dq^2$ is the total branching fraction in the decay $\bdtau$ with
$\Gamma$ as the total decay width. $N_{sig}$ is the total number of events. As given in eq.(\ref{eq5}), 
the errors are also related to the production cross section $\sigma_P$ ( = $\sigma_{\bdtau}/{\cal B}(\bdtau)$), 
and the effective luminosity ${\cal L}_{eff} = {\cal L}_{int} \epsilon_s$, where 
${\cal L}_{int}$ and $\epsilon_s$ are the integrated luminosity and reconstruction efficiency respectively
\footnote{As we know that the cross section $\sigma_{a\to b} = \sigma_a \Gamma_b/\Gamma$, therefore,
we can define $\sigma_{\bdtau} = \sigma_P {\cal B}(\bdtau)$, where $\sigma_P$ is the $B{\bar B}$ production cross section.
If we redefine our observable as $\sigma_{\bdtau}$ than the errors in $C_i$ can be written as
\bea
\delta C_i &=&\sqrt{\frac{X_{ii}' \sigma_{\bdtau}}{ N_{sig}}} =
\sqrt{\frac{X_{ii}'}{{\cal L}_{eff}}} = \sqrt{\frac{X_{ii}}{\sigma_P {\cal L}_{eff}}} \nonumber \\
 &=& \sqrt{\frac{X_{ii} {{\cal B}(\bdtau)}^{exp}}{N_{sig}}} , \nn
\eea
since $X_{ii}' = X_{ii}/\sigma_P$.} .     
The above-mentioned method, and the equations like (\ref{eq4}) and (\ref{eq5}),
can be generalised for any other observables in $\bdtau$ decay. 

Since the data is consistent with the SM, if there is NP in $\bdtau$ decays, the effect is expected to 
be small compared to their SM counterpart. The earlier model independent analysis \cite{Sakaki:2014}, 
which is based on data by BABAR \cite{Lees:2013uzd}, shows that zero value of the new Wilson coefficients 
are consistent with the data. Therefore, we choose our starting point as $C_W = 0$ and find out errors in the 
extraction of those coefficients around that point.
In addition to that, we assume that the error on $C_i$ could be captured sufficiently well by just the leading-order terms. 

We focus on the following observables:
%\begin{widetext}
\begin{itemize}

\item{ The branching fractions, obtained by integrating the differential branching fractions over
the full $q^2$ region, normalised by the full $q^2$ integrated branching fraction ${\cal B}_{\ell} = {\cal B}(\bdell)$.
\be
R(D^{(\ast)}) = \int dq^2 R_{D^{(\ast)}}(q^2), 
%= \frac{1}{{\cal B}(\bdell)}\int dq^2\frac{d{\cal B}(\bdtau)}{dq^2},
\label{eq7}
\ee
%\end{widetext}
with
\be
R_{D^{(\ast)}}(q^2) = \frac{1}{{\cal B}_{\ell}} \frac{d{\cal B}(\bdtau)}{dq^2}.
\label{eq8}
\ee}
\end{itemize}
\begin{itemize}
\item{$\tau$ polarisation asymmetry which we defined as $P_{\tau}^{R^{(*)}}(q^2) = P_\tau(q^2) R_{D^{(\ast)}}(q^2)$,
where
%\begin{widetext}
\be
P_{\tau}(q^2) = \frac{d\Gamma_{\lambda=1/2}/dq^2 - d\Gamma_{\lambda=-1/2}/dq^2}
{d\Gamma_{\lambda=1/2}/dq^2 + d\Gamma_{\lambda=-1/2}/dq^2}.
\label{eq9}
\ee}
\end{itemize}

\begin{itemize} 
\item{$\tau$ forward-backward asymmetry ${\cal A}_{FB}^{R^{(*)}}(q^2) = {\cal A}_{FB}(q^2) R_{D^{(\ast)}}(q^2)$, 
where
%\begin{widetext}
\bea
{\cal A}_{FB}(q^2) &=&\frac{\int_{0}^{1} \frac{d\Gamma}{dq^2 d\cos\theta} d\cos\theta-\int_{-1}^{0}
 \frac{d\Gamma}{dq^2 d\cos\theta}d\cos\theta}
    {\int_{-1}^{1} \frac{d\Gamma}{dq^2 d\cos\theta} d\cos\theta} \nonumber \\ 
    &=& \frac{b_{\theta}(q^2)}{d \Gamma / dq^2},
\label{eq10}
\eea
where $\theta$ is the angle that $\tau$ makes with the $\bar B$ in the rest frame of $\tau{\bar\nu}$. The expressions for
$b_{\theta}(q^2)$ are given in \cite{Sakaki:2013}.}
\end{itemize}

\begin{itemize}
\item{$D^{\ast}$ longitudinal polarisation asymmetry $P_{D^{\ast}}^R (q^2)=P_{D^{\ast}}(q^2) R_{D^{\ast}}(q^2) $,
 where
\be
P_{D^{\ast}}(q^2)  = \frac{\frac{d\Gamma}{dq^2}(\lambda_{D^{\ast}=0})}{\frac{d\Gamma}{dq^2}(\lambda_{D^{\ast} = 0})
+ \frac{d\Gamma}{dq^2}(\lambda_{D^{\ast} = 1}) + \frac{d\Gamma}{dq^2}(\lambda_{D^{\ast} =-1})}.
\label{eq12}
\ee}
\end{itemize}

In the above definitions, the detailed expression for $d\Gamma/{dq^2}$ are taken from ref.\cite{Sakaki:2013}.
For forward backward asymmetries and the
$D^*$ polarisation, the $C_i$s and the corresponding $f_i(q^2)$s are given in the Tables \ref{cisasy}, and in the Appendix 
\ref{fisasy}. 

\begin{table*}[!htbp]
\begin{center}
\begin{tabular}{|c|c|c|c|}
\hline
\cline{2-4}
\backslashbox{$C_i$}{Obs} & ${\cal A}_{FB}^R(q^2)$ &  ${\cal A}_{FB}^{R^{\ast}}(q^2)$  & $P_{D^{\ast}}(q^2)$ \\
\hline
$C_1$ & $|1 + C_{V_1} + C_{V_2}|^2$ &  $|1 + C_{V_1}|^2 - |C_{V_2}|^2$ & $|1 + C_{V_1} - C_{V_2}|^2$ \\
\hline
$C_2$ &  $Re[(1 + C_{V_1} + C_{V_2})(C_{S_1}^{\ast} + C_{S_2}^{\ast})]$ & $|1 + C_{V_1} - C_{V_2}|^2$  & $|C_{S_1} - C_{S_2}|^2$ \\
\hline
$C_3$ & $Re[(1 + C_{V_1} + C_{V_2})C_T^{\ast}]$  &  $|C_T|^2$ & $|C_T|^2$    \\
\hline
$C_4$ & $Re[(C_{S_1} + C_{S_2})C_T^{\ast}]$ &  $Re[(1 + C_{V_1} - C_{V_2})(C_{S_1}^{\ast} - C_{S_2}^{\ast})]$ &
                                            $Re[(1 + C_{V_1} - C_{V_2})(C_{S_1}^{\ast} - C_{S_2}^{\ast})]$\\
\hline
$C_5$ & --  &  $Re[(1 + C_{V_1})C_T^{\ast}]$   &  $Re[(1 + C_{V_1} - C_{V_2})C_T^{\ast}]$ \\
\hline
$C_6$ &   --     &   $Re[C_{V_2}C_T^{\ast}]$              &   --   \\
\hline
$C_7$ &   --   &  $Re[(C_{S_1} - C_{S_2})C_T^{\ast}]$        &  --     \\
\hline
\end{tabular}
\end{center}
\caption{Expressions of $C_i$s for different observables.}
\label{cisasy}
%\end{center}
\end{table*}

All the above-mentioned observables are expected to be measured with good statistics in future experiments 
like Belle-II and LHCb. The corresponding errors on $C_i$ can be obtained using the following relation
\be
|\delta C_i| = \sqrt{\frac{X_{ii}^{\ell}}{{\cal B}_{\ell}~\sigma_P~ {\cal L}_{eff}}},
\ee 
where $X_{ii}^\ell = X_{ii} {\cal B}_{\ell}$.

\section{Analysis}
There are varieties of NP models that can contribute to $\bdtau$, and the characteristics of those
models could be very different. For example, two higgs doublet model (2HDM)
has only scalar-type interactions, new gauge boson $Z'$ and $W'$ take part only in vector-type
interactions, the model with leptoquarks has both the scalar or vector-type interactions \cite{Davidson:1993}
\footnote{Although the model with scalar-type interactions may also contribute to tensor Wilson coefficients by Fierz
reordering.}, the extra dimensional models have tensor interaction in addition to scalar or vector-type interactions
\cite{Ponton:2012}. 

In our analysis of the decay $\bd0tau$, it will be hard to estimate the uncertainties in the extractions of
$C_{V_1}$ and $C_{V_2}$, because they cannot be singled out from their SM counterpart (same $f_i$s). 
The similar argument holds for the decay $\bdasttau$, however, in this decay we can estimate the error in the extraction 
of $Re(C_{V_2})$, see for instance Tables \ref{cird} and \ref{fisrdast} where $f_2$ associated with $C_2$ is 
different from $f_1$ associated with $C_1$. In Table \ref{casebd}, we list a few interesting cases of NP relevant 
for the observables in $\bd0tau$. In many cases, we assume $C_V = C_{V_1} + C_{V_2} =0$, however, 
the assumption $C_V \ne 0$ will lead to the same set of parameters that has to be simultaneously extracted,
if it is assumed that $C_V << 1$. Under such 
conditions $C_1$ can be treated as the Wilson coefficient of the vector operator.  
The different NP cases related to the observables in $\bdasttau$ are given 
in Tables \ref{casesrdast}, \ref{casesafbast} and \ref{casesdastpol} respectively.
In most cases, we assume $C_{V_1} = 0$, though the same set of parameters can be obtained without 
this assumption if $C_{V_1} << 1$. 
For $\tau$ forward-backward asymmetry in $\bdasttau$, we discuss mostly the cases 
with $C_{V_2} = 0$.  In such cases, $C_1 = C_2$ , and therefore we need to merge $f_1(q^2)$ and $f_2(q^2)$ into 
$f(q^2)( = f_1(q^2) +  f_2(q^2))$ for the analysis. In all the other cases, when $C_{V_2} \neq 0$, the 
extracted uncertainties are large. We will discuss only one such interesting case.

\begin{table}[!htbp]
\begin{center}
%\subfloat[Cases relevant for $R_{D}(q^2)$ and $P_{\tau}^R(q^2)$]
\begin{tabular}{| c | c |}
\hline
Cases & Assumptions \\
\hline
$a$ & $C_i \neq 0$, $i=1,..5$ \\
\hline
$b$ & $Re(C_T) = 0$ \\
\hline
$c$ & $Re(C_{S})=0$  \\
\hline
$d$ & $Re(C_{S})=0$ and $Re(C_{T}) =0$  \\
\hline
$e$ & $C_{T} = 0$ \\
\hline
$f$ & $C_{S} = 0$ \\
\hline
\end{tabular}
\end{center}
\caption{Cases relevant for $R_{D}(q^2)$, $P_{\tau}^R(q^2)$ and ${\cal A}_{FB}^R(q^2)$. Here $C_S = C_{S_1} + C_{S_2}$, 
and in all the cases $C_{V_1} + C_{V_2} =0$.}
%\end{center}
\label{casebd}
\end{table}

\begin{table}[htbp!]
\begin{center}
\begin{tabular}{ | c | c |}
\hline
 Cases & Assumptions \\
\hline
$a^{\ast}$ &  $C_S = 0$\\
\hline
$b^{\ast}$& $C_{V_2}=0$\\
\hline
$c^{\ast}$ &  $C_T =0$ \\
\hline
$d^{\ast}$ & $Re(C_S) =0$,\ \ $ C_{V_2} = 0$  \\
\hline
$e^{\ast}$ & $Re(C_S) =0$,\ \ $C_{T}= 0$,\ \ $Im(C_{V_2}) = 0$\\
\hline
$f^{\ast}$ & $C_{S} =0$,\ \ $Re(C_{T})= 0 $,\ \ $ Im(C_{V_2})=0 $ \\
\hline
$g^{\ast}$ & $C_{V_2} =0$, \ \ $Re(C_T) =0$, \ \ $Re(C_S) = 0$ \\
\hline
$h^{\ast}$ & $C_{V_2} =0$,\ \ $C_T = 0$\\
\hline
$i^{\ast}$ & $C_{V_2}=0$,\ \  $C_S=0$ \\
\hline
$j^{\ast}$ & $Re(C_S)=0$,\ \ $Re(C_T)=0$,\ \ $Im(C_{V_2}) =0$ \\
\hline
$k^{\ast}$ & $C_S=0$,\ \ $C_T=0$ \\
\hline
\end{tabular}
%\caption{Different cases related to $R_{D^*}(q^2)$ and $P_{\tau}^{R^*}(q^2)$ in $\bdasttau$. Here $C_S = C_{S_1} - C_{S_2}$}
%\label{casesrdast}
\end{center}
\caption{Different cases related to $R_{D^*}(q^2)$ and $P_{\tau}^{R^*}(q^2)$ in $\bdasttau$. Here $C_S = C_{S_1} - C_{S_2}$,
 and in all the cases $C_{V_1} = 0$.}
\label{casesrdast}
\end{table}

\vskip 2pt
\begin{table}[!htbp]
\begin{center}
%\subfloat[NP cases relevant in ${\cal A}_{FB}^{R^*}(q^2)$ with $C_{V_2} = 0$.]
\begin{tabular}{| c | c |}
\hline
Cases & Assumptions  \\
\hline 
$1^{\ast}$ & $C_S=0$ \\ 
\hline
$2^{\ast}$ & $C_{V_2}=0$,\ \ $Re(C_T) = 0$ \\
\hline
$3^{\ast}$ & $C_{V_2}=0$,\ \  $Re(C_S)= 0 $ \\
\hline
$4^{\ast}$ & $C_{V_2}=0$, \ \  $Re(C_T) =0$, $Im(C_S) = 0 $\\
\hline
$5^{\ast}$ & $C_{V_2}=0$,\ \ $ C_S = 0 $ \\
\hline
$6^{\ast}$ &$C_{V_2}=0$,\ \  $Re(C_S) =0$, $ Re(C_T) = 0 $ \\
\hline
$7^{\ast}$ &$C_{V_2}=0$,\ \ $ C_T = 0$ \\
\hline
\end{tabular}
\end{center}
\caption{Cases relevant in ${\cal A}_{FB}^{R^*}(q^2)$ with $C_{V_1} = 0$.}
\label{casesafbast}
\end{table}

\vskip 2pt
\begin{table}[!htbp]
\begin{center}
%\subfloat[NP cases relevant in $D^{\ast}$ polarisation asymmetry. Here, $C_V = C_{V_1} - C_{V_2}$. ]
\begin{tabular}{| c | c |}
\hline
Cases & Assumptions  \\
\hline
$A$ & $C_i \neq 0$, $i=1,..,5$  \\
\hline
$B$ & $Re(C_T) = 0$  \\
\hline
$C$ & $Re(C_S) = 0$  \\
\hline
$D$ & $Re(C_S) =0$,\ \ $Re(C_T) = 0$  \\
\hline
$E$ & $ C_{T} = 0 $ \\
\hline
$F$ & $ C_{S} = 0 $ \\
\hline
 $G$ & $Re(C_S) =0$,\ \ $C_T = 0$ \\
\hline
$H$ & $Re(C_T) =0$,\ \  $C_S = 0$ \\
\hline
\end{tabular}
\end{center}
\caption{NP cases relevant in $D^{\ast}$ polarisation asymmetry. Here, $C_V = C_{V_1} - C_{V_2} = 0$.}
\label{casesdastpol}
\end{table}

The numerical values of all the relevant parameters, like the form-factors, various masses and lifetimes are 
taken from ref. \cite{formfactors}, and for the analysis we choose the central values of all 
the form-factors. The errors of the form-factors are considered while we estimate the additional errors 
on the extracted coefficients. We choose as benchmark values ${\cal B}(B\to D \ell\nu) = 2.32$\%, 
${\cal B}(\bdastell) = 5.31$\%, $\sigma_P = 1105.63$ {\it pb}, and ${\cal L}_{eff} =1 fb^{-1}$.

%\begingroup
%\squeezetable
\begin{table}[!htbp]
\begin{center}
{\begin{tabular}{| c | c | c | c | c | c |}
\hline
Decay & \multicolumn{4}{c|}{$\bd0tau$} \\
\hline
Cases & \multicolumn{2}{ c | }{${\it b}$} & \multicolumn{2}{ c | }{${\it c}$}  \\
\cline{1-5}
\backslashbox{$|\delta C_i|$}{Obs.} & $R(D)$ & $P^R_\tau$ & $R(D)$ & $P^R_\tau$  \\
\hline
$\delta C_1$ & 0.60 & 0.37 & 0.60 & 0.37  \\
\hline
% \hline
$\delta|C_S|^2$ & 1.03 & 0.04 & 0.13 & 0.08 \\
\hline
$\delta|C_T|^2$ & 0.62 & 0.70 & 1.12 & 0.72  \\
\hline
$\delta(Re(C_S))$ & 1.31 & 0.06 & - & - \\
%\hline
%$\delta(Im(C_S))^2$ & - & - & 0.13 & 0.08 & - & - & - & - & 1.00 & 1.73 & - & - & - & - \\
\hline
$\delta(Re(C_T))$ & - & - & 1.15 & 0.12  \\
%\hline
%$\delta(Im(C_T))^2$ & 0.62 & 0.70 & - & - & 0.26 & 0.09 & 0.08 & 0.02 & - & -& - & - & - & -\\
\hline
% $\delta(Re(C_{V_2}))$ & \multicolumn{8}{c|}{N.A.} & 46.59 & 30.08 & - & - & - &- & - & - & - & -\\
% \hline
\end{tabular}}
\end{center}
\caption{Numerical values of the 1$\sigma$ error on $C_i$s extracted from $R(D)$ and $P_{\tau}^R$. For the cases
$Re(C_i) = 0$, $\delta |C_i|^2 = \delta (Im^2(C_i))$.}
\label{tablerd}
\end{table}

In Table \ref{tablerd}, we list our main results of the uncertainties in $C_i$ extracted from
the analysis of the $R(D)$ and $P_{\tau}^R$ corresponding to different cases listed in Table \ref{casebd},
while those for $R(D^*)$ and $P_{\tau}^{R^*}$, corresponding to the cases listed in Table \ref{casesrdast},
are given in Table \ref{tablerdast}. For a given case, we estimate the statistical significance of the simultaneous 
extraction of $C_i$s. The numerical values are given only for parameters relevant to a particular case, 
while the rest are set to zero. 

\begingroup
\begin{table*}[!htbp]
\begin{center}
\begin{tabular}{| c | c | c |c  | c |c | c | c | c| c | c |}
\hline
Decay Modes & \multicolumn{10}{|c|}{$\bdasttau$} \\
\hline
Cases & \multicolumn{2}{c |}{$a^*$} & \multicolumn{2}{c |}{$b^*$} & \multicolumn{2}{c |}{$c^*$} & 
  \multicolumn{2}{c|}{${\it d^*}$} & \multicolumn{2}{c|}{${\it e^*}$} \\
\cline{1-11}
\backslashbox{Params.}{Obs.} & $R(D^*)$ & $P^{R^*}_\tau$ & $R(D^*)$ & $P^{R^*}_\tau$ & $R(D^*)$ & $P^{R^*}_\tau$ & 
 $R(D^*)$ & $P^{R^*}_\tau$ & $R(D^*)$ & $P^{R^*}_\tau$ \\
\hline
$\delta C_1$ & 7.22 & 13.70 & 289.17 & 116.82 & 28.44 & 14.08 & 2.01 & 0.65 & 1.28 & 1.25 \\
\hline
% \hline
$\delta|C_S|^2$ & - & - & 629.08  & 204.37  & 56.25 & 29.65 & 1.00 & 1.73 & 3.68 & 1.83  \\
\hline
$\delta|C_T|^2$ & 4.30 & 1.96 & 11.86 & 4.62 & - & - & 0.03 & 0.04 & - & - \\
\hline
$\delta(Re(C_S))$ & - & - & 529.3 & 191.49 & 6.81 & 2.20 & - & - & - & -     \\
%\hline
%$\delta(Im(C_S))^2$ & - & - & 0.13 & 0.08 & - & - & - & - & 1.00 & 1.73 & - & - & - & - \\
\hline
$\delta(Re(C_T))$  & 28.27 & 36.92 & 36.71 & 45.37 & - & - & 0.35 & 0.24 &-&-  \\
%\hline
%$\delta(Im(C_T))^2$ & 0.62 & 0.70 & - & - & 0.26 & 0.09 & 0.08 & 0.02 & - & -& - & - & - & -\\
\hline
$\delta(Re(C_{V_2}))$ & 47.10 & 18.51 & - & - & 14.21 & 7.03 & - & - & 0.63 & 0.63   \\
\hline
$\delta Re[C_{V_2} C_T^*]$ & 15.70 & 17.19 & - & - & -  &  - & - & - & - & -  \\
\hline
\end{tabular}
\end{center}
\caption{Numerical values of the 1$\sigma$ error on $C_i$s extracted from $R(D^*)$ and $P_{\tau}^{R^*}$.
For the cases $Re(C_i) = 0$, $\delta |C_i|^2 = \delta (Im^2 (C_i))$.}
\label{tablerdast}
\end{table*}

\begin{figure*}[!htbp]
\subfloat[Case ${\it d}$ ($R(D)$)]
{\includegraphics[scale=0.23]{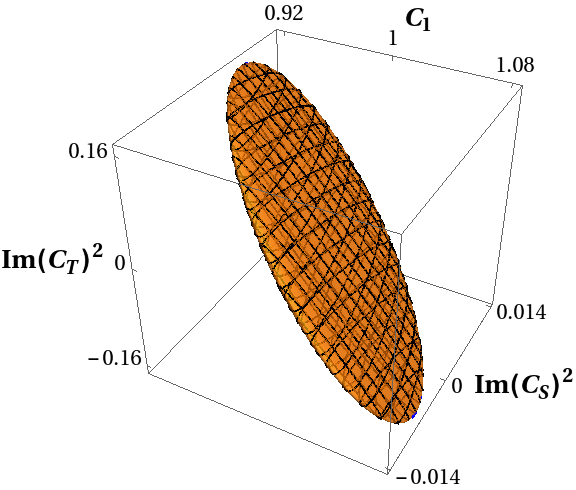}\label{rdc}}~~~~ 
\subfloat[Case ${\it e}$ ($R(D)$)]
{\includegraphics[scale=0.23]{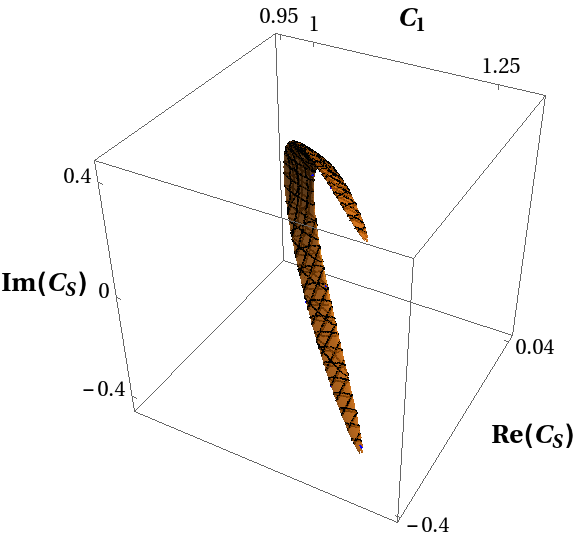}\label{rdd}}~~~
\subfloat[Case ${\it f}$ ($R(D)$)]
{\includegraphics[scale=0.23]{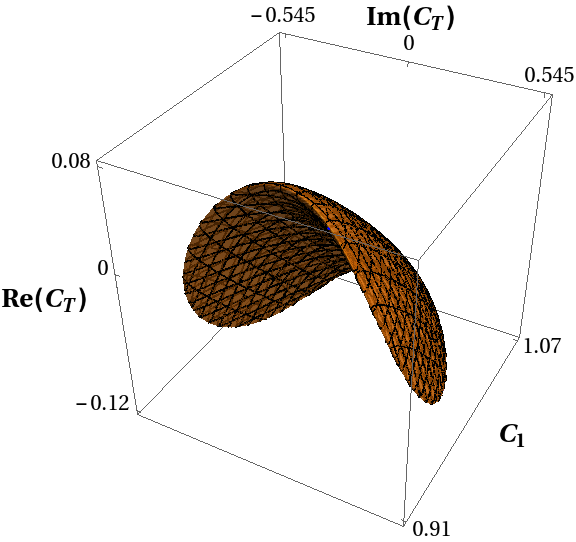}\label{rdf}}\\
\subfloat[Case ${\it d}$ ($P_\tau^R$)]
{\includegraphics[scale=0.25]{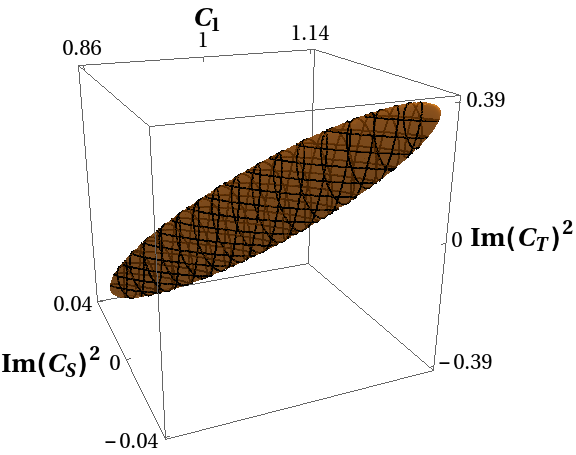}\label{ptauc}}~~~~
\subfloat[Case ${\it e}$ ($P_\tau^R$)]
{\includegraphics[scale=0.23]{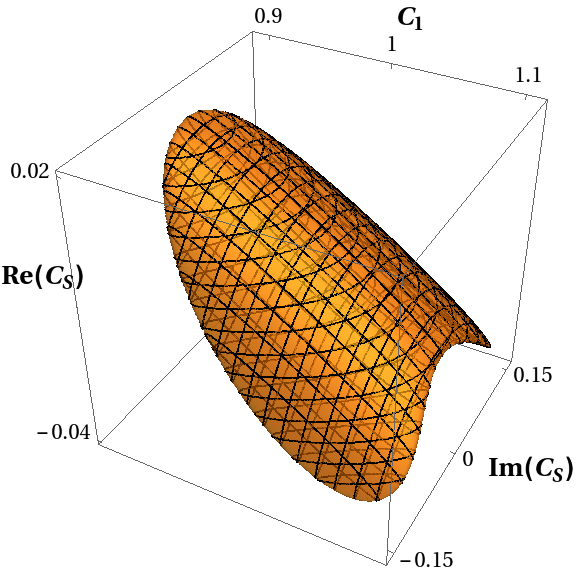}\label{ptaud}}~~~~ 
\subfloat[Case ${\it f}$ ($P_\tau^R$)]
{\includegraphics[scale=0.23]{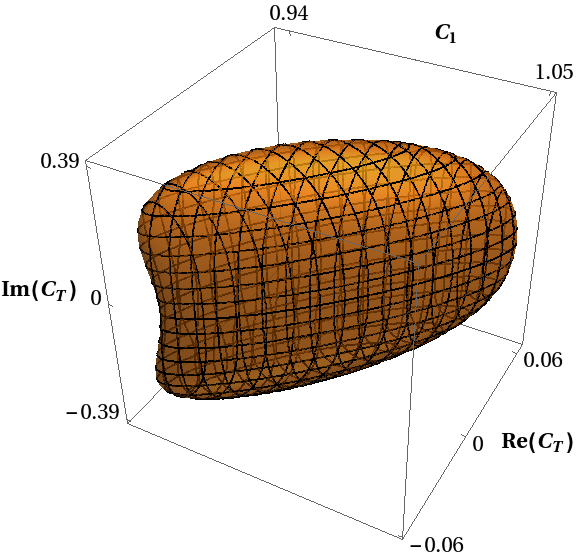}\label{ptauf}}\\
\caption{Surfaces of constant $\chi^2 = 1$ for a few selected cases of different observables in $\bd0tau$.}
\label{figcasesbd}
\end{figure*}

\begin{figure*}[!htbp]
\subfloat[Case ${\it f^*}$ ($R(D^*)$) ]
{\includegraphics[scale=0.21]{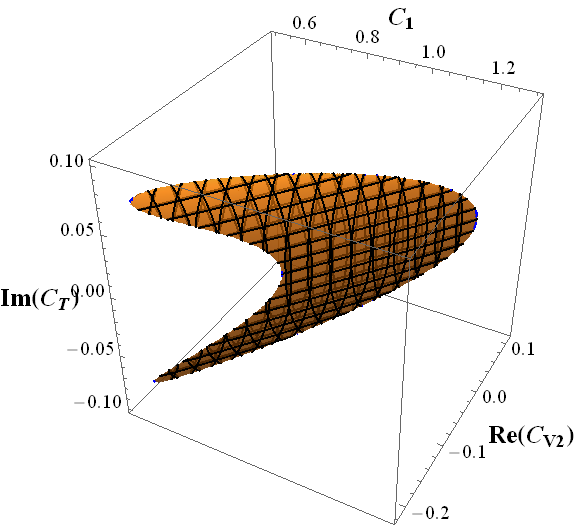}\label{rdstf}}~~~
\subfloat[Case ${\it g^*}$ ($R(D^*)$) ]
{\includegraphics[scale=0.21]{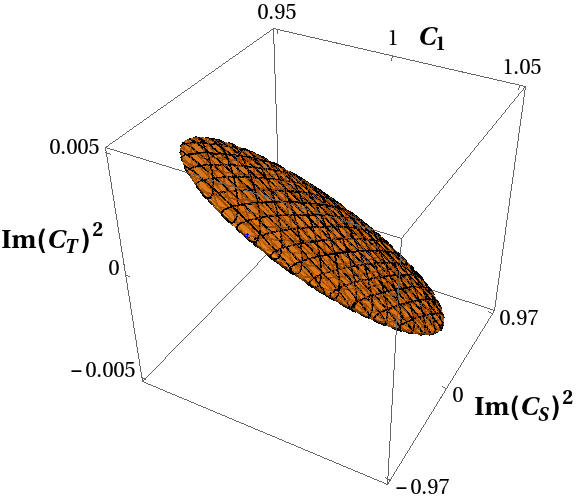}\label{rdstg}}~~~
\subfloat[Case ${\it h^*}$ ($R(D^*)$) ]
{\includegraphics[scale=0.21]{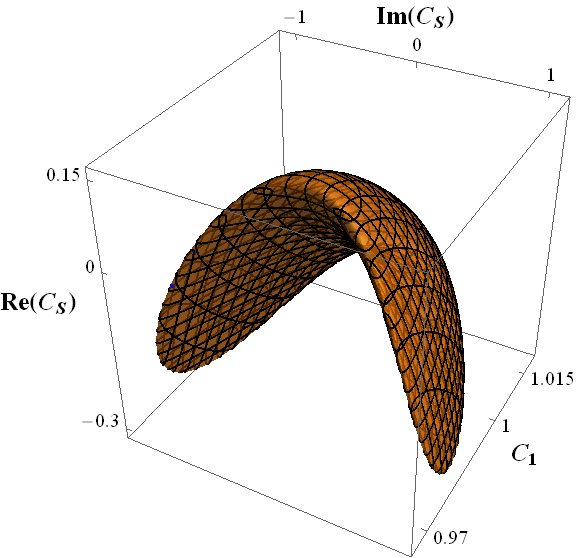}\label{rdsth}}~~~~
\subfloat[Case ${\it i^*}$ ($R(D^*)$) ]
{\includegraphics[scale=0.21]{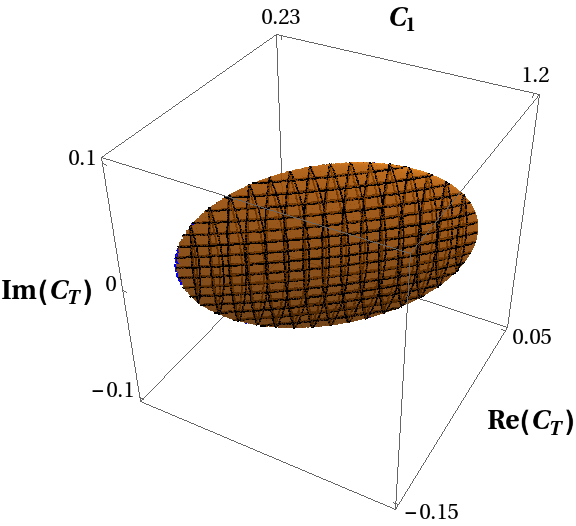}\label{rdsti}}\\
\subfloat[Case ${\it f^*}$ ($P_\tau^{R^*}$) ]
{\includegraphics[scale=0.21]{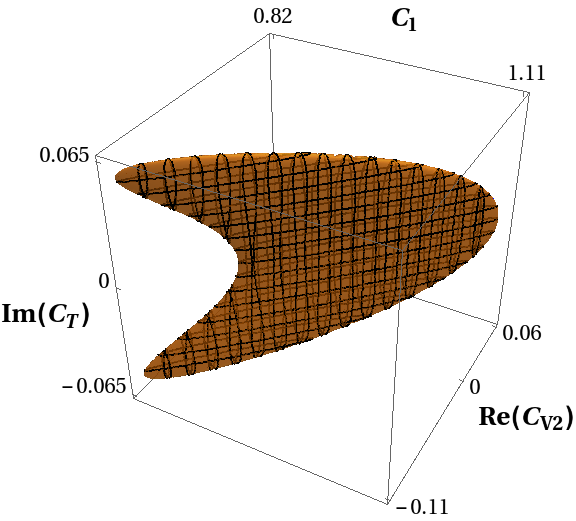}\label{ptaustf}}~~~
\subfloat[Case ${\it g^*}$ ($P_\tau^{R^*}$) ]
{\includegraphics[scale=0.21]{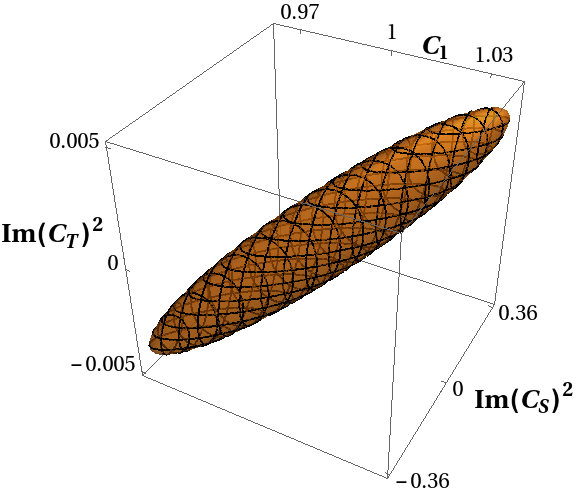}\label{ptaustg}}~~~
\subfloat[Case ${\it h^*}$ ($P_\tau^{R^*}$) ]
{\includegraphics[scale=0.21]{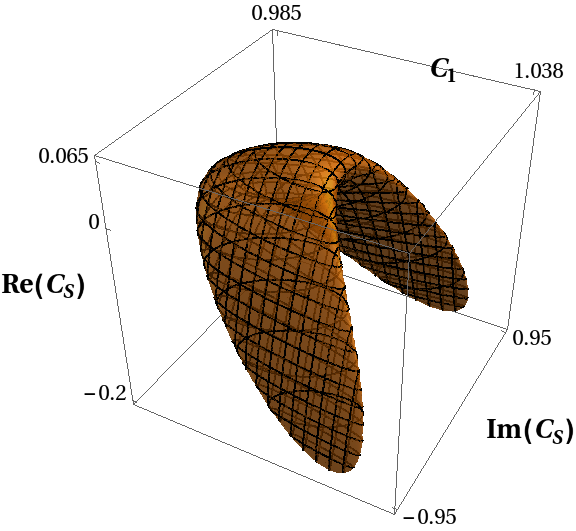}\label{ptausth}}~~~
\subfloat[Case ${\it i^*}$ ($P_\tau^{R^*}$) ]
{\includegraphics[scale=0.21]{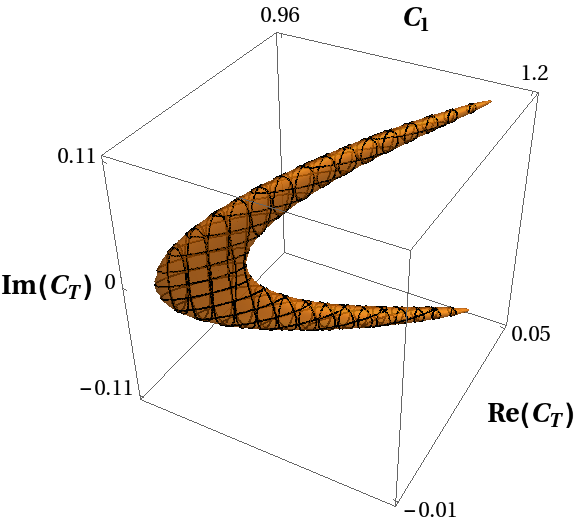}\label{ptausti}}~~~
\caption{Surfaces of constant $\chi^2 = 1$ for a few selected cases of different observables shown in $\bdasttau$.}
\label{figcasesbdst}
\end{figure*}
%\end{widetext}

In some more simplified cases, where the number of non-zero NP parameters are less, we compute the $\chi^2$, 
which is defined as
\begin{align}
\nn \chi^2 &= \sum_{i,j} (C_i - C_i^0) (C_j - C_j^0) V_{ij}^{-1},~ \\
\text{where, }~~~V_{ij} &= \frac{X_{ij}}{{\cal B}_{\ell}~\sigma_P~{\cal L}_{eff}}.
\end{align}
The $C_i^0$s are the seed values, which can be considered as model inputs; as discussed earlier, we choose $C_i^0=0$,
for $i\neq 1$, and $C_1^0 = 1$. The $\chi^2 = 1$ surfaces are perfect ellipsoids in $C_i$ basis, and they indicate the
$\pm 1\sigma$ errors in the determination of $C_i$s. The constant $\chi^2 = 1$ surfaces are shown in Figs.
\ref{figcasesbd}, and \ref{figcasesbdst}. The largest and the smallest values in the figures represent 
$\pm 1 \sigma$ errors of corresponding parameters.

\begin{table}[!htbp]
\begin{center}
%\subfloat[Numerical values of $|\delta C_i|$ extracted from ${\cal A}_{FB}^{R}(q^2)$.]
{\begin{tabular}{| c | c | c |c |c | c |}
\hline
%\multicolumn{6}{|c|}{$\sqrt{X_{ii}}$} \\
%\cline{1-6}
\backslashbox{$|\delta C_i|$}{Cases} & b & c  & d &  e  & f     \\
\hline
 $|\delta C_1|$  & 0.27 & 0.27 & 0.04 & 0.07 & 0.07 \\
\hline
$|\delta Re (C_S)|$ &  0.40 & -   & -  & 0.05  &  -     \\
\hline
$|\delta Re (C_T)|$  & -      & 0.30 & - &   -    &  0.04  \\
%\hline
%$|\delta(Re(C_S) Re({C_T}^*))|$  & - & - &  -    &   -   &  - \\
\hline
$|\delta(Im(C_S) Im({C_T}))|$  & 0.11 & 0.11 & 0.01 &  -    &   -  \\
\hline
\end{tabular}}
\end{center}
\caption{Numerical values of $|\delta C_i|$ extracted from the $\tau$ forward-backward asymmetry in $\bd0tau$.}
\label{afbbd}
\end{table}

\subsection{Discussions}
For case {\it a}, the uncertainties obtained from the simultaneous extraction of all the Wilson coefficients  
from the observable $R(D)$ shows that \footnote{The results corresponding to the case {\it a} are not shown in the table,
because the extracted uncertainties are very large $(>>1)$. } 
\be
\frac{|\delta C_4|}{|\delta C_5|} \approx \frac{|\delta Re(C_S)|}{|\delta Re(C_T)|} \sim 1, \, \, \, \, \, \, 
\frac{\delta |C_S|^2}{\delta |C_T|^2} \sim 2,
\ee
which shows that $R(D)$ is equally sensitive to the real part of $C_S$ and $C_T$. The above result 
does not allow a direct comparison between the sensitivities to the imaginary part of the coefficients.   
The results obtained for all the other cases are shown in Tab. \ref{tablerd}, and in 
Figs.\ref{rdc}, \ref{rdd} and \ref{rdf}. We note that if the Wilson coefficients are purely imaginary then 
$R(D)$ is more sensitive to $Im(C_S)$ compared to $Im(C_T)$. Also, it is important to note that 
this observable is more sensitive to the real part of the coefficients than the imaginary part. 
Comparing all the different cases considered for $R(D)$, it would be difficult to comment on the overall 
sensitivity of this observable to a particular type of NP interaction. However, in the next section we will 
see that there are distinct regions of $q^2$ which are sensitive to either scalar or tensor type interactions. 

On the other hand, the analysis of $P_{\tau}^R(q^2)$ for case {\it a} gives    
\be
\frac{|\delta C_4|}{|\delta C_5|} \approx \frac{|\delta Re(C_S)|}{|\delta Re(C_T)|} \sim 0.5, \, \, \, \, \, \, 
\frac{\delta |C_S|^2}{\delta |C_T|^2} \sim 2,
\ee
which shows an improvement in sensitivity to $Re(C_S)$ compared to $Re(C_T)$. 
The results obtained from all the other relevant cases are shown in Table \ref{tablerd}, and in 
Figs. \ref{ptauc}, \ref{ptaud}, and \ref{ptauf}, which allow a case by case comparison between the 
results obtained from $R(D)$ and $P_{\tau}^R$. Interestingly, the extracted uncertainties are less compared 
to that extracted in $R(D)$. The sensitivity of this observable to tensor interaction is a little less,
compared to scalar interaction, but it can be extracted with uncertainties less than 1. 

As shown in Table \ref{tablerdast}, and in Fig. \ref{figcasesbdst}, 
the observables like $R(D^*)$ and $P_{\tau}^{R^*}$ are more sensitive to $|C_T|^2$ compared to 
any other new Wilson coefficients, almost in all the cases $|C_T|^2$ and $Re(C_T)$ can be 
extracted with small uncertainties. A case by case comparison shows that
the above observables are more sensitive to $C_T$ than $C_S$,  and $\delta Re(C_T) < \delta Re(C_{V_2})$ but
they are of same order. Also, when $C_T$ is purely imaginary, we find $\delta Im(C_T) \approx \delta Re(C_{V_2})$, though  
$P_{\tau}^{R^*}$ have little better sensitivity to $Im(C_T)$. Therefore, these observables alone won't allow 
us to distinguish the contributions from right handed vector current to that of a tensor current. 
However, in the next section we will see that a bin by bin analysis of the $q^2$ distribution of 
the differential decay rate allows to discriminate the effects of these interactions.  
 
However, when $C_T = 0$, both the observables are almost equally sensitive, though 
$\delta Re (C_S) < \delta Re(C_{V_2})$,  to the real part of the vector and scalar Wilson coefficients.
In case $C_S$ is purely imaginary, $\delta Re(C_{V_2}) << \delta Im(C_S)$ i.e the observables are less sensitive 
to the imaginary part of $C_S$ compared to the real parts of $C_{V_2}$ and $C_S$. 
Again, we note that the extracted errors on $C_i$s from $P_{\tau}^{R^*}$ are smaller than those in $R(D^*)$.  

\begin{table}[!htbp]
\begin{center} 
{\begin{tabular}{| c | c | c | c | c | c | c |}
\hline
% \multicolumn{7}{|c||}{$\sqrt{X_{ii}}$} \\
%\cline{1-7}
\backslashbox{$|\delta C_i|$}{Cases}& $2^*$ & $3^*$ & $4^*$ & $5^*$ & $6^*$ & $7^*$ \\
\hline
$|\delta C_1|$ & 1.30 & 2.41 & 0.40 & 0.27 & 0.52 & 0.04 \\
\hline
$\delta |C_T|^2$ & 0.12 & 0.08 & 0.04 & 0.04 & 0.03 & -\\
\hline
$|\delta Re (C_S)|$ & 16.30 & - & 1.72 & - & - & 0.02 \\
\hline
$|\delta Re (C_T)| $ & - & 1.06 & - & 0.12 & - & -\\
\hline
$|\delta(Im(C_S) Im({C_T}^*))|$ & 2.48 & 2.41 & - & - & 0.26 & -\\
\hline
\end{tabular}}\\
%\label{afb}
\end{center}
\caption{Numerical values of $|\delta C_i|$ extracted from $\tau$ forward backward asymmetry in $\bdasttau$.}
\label{afbst}
%\end{center}
\end{table}

The results of the analysis of forward-backward asymmetries and $D^*$ polarisation are given 
in Tables \ref{afbbd}, \ref{afbst}, and \ref{dastpol} respectively. The forward-backward asymmetry
in $\bd0tau$ is equally sensitive to the scalar and tensor type interactions. 

 For case $1^*$ in $A_{FB}^{R^*}$, we find 
\be
\frac{\delta C_1}{\delta |C_T|^2} \approx 1, \ \ \ \  \frac{\delta C_2}{\delta |C_T|^2} \approx 24,
\ee
and 
\be
\frac{\delta C_2}{\delta Re(C_T)} \approx 12 ,
\ee
where $C_1$ and $C_2$ are the functions of the Wilson coefficients of the vector operators. The approximate forms
are given by 
\be
C_1 \approx 1 + 2 Re(C_{V_1}), \  C_2 \approx 1 + 2 Re(C_{V_1}) - 2 Re(C_{V_2}).
\ee
It indicates that the $\tau$ forward-backward asymmetry in $\bdasttau$ is more sensitive to tensor Wilson coefficients
than to a vector, in particular to $C_{V_2}$. In order to understand it better, we define 
\be
C_{12} = C_1 - C_2 \approx 2 Re(C_{V_2}).
\ee
Therefore, a simple calculation shows that 
\be
\frac{\delta Re(C_{V_2})}{\delta Re(C_T)} =\frac{1}{2} \frac{\delta C_{12}}{ \delta Re(C_T)} \approx 6.  
\ee
In all the other cases with $C_{V_2}=0$, the $A_{FB}$ in $\bdasttau$ is more sensitive to the tensor interaction compared 
to the scalar. On the other hand the $D^*$ polarisation is equally sensitive to the scalar and tensor interactions.
Therefore, if future data shows large deviations from the SM predictions in all the observables 
like $R(D^*)$, $A_{FB}^*$, and $D^*$ polarisation, that can be thought of as an indication of 
the presence of a new tensor type interaction. On other hand, if a deviation is only in $R(D^*)$ and not in the others, 
that could be an indication of a new vector interaction.

\begin{table}[htbp]
\begin{center}
%\subfloat[Numerical values of $\sqrt{X_{ii}}$.]
%The errors in $C_i$ would be $\delta C_i = \sqrt{X_{ii} (P_{D^*}^R)^{exp}/N}$]
{\begin{tabular}{| c | c | c | c | c | c | c | c | c|}
\hline
%\multicolumn{9}{|c|}{$\sqrt{X_{ii}}$  } \\
%\cline{1-9}
\backslashbox{$|\delta C_i|$}{Cases} & $A$ & $B$ & $C$ & $D$ & $E$ & $F$ & $G$ & $H$ \\
\hline
$|\delta C_1|$ & 3.58 & 0.74 & 1.41 & 0.12 & 0.05 & 1.41 & 0.05 & 0.1 \\
\hline
$\delta |C_S|^2 $ & 16.63 & 3.39 & 0.55 & 0.55 & 0.77 & - & 0.54 & -\\
\hline
$\delta |C_T|^2$ & 2.53 & 0.09 & 0.41 & 0.01 & - & 0.41 & - & 0.01 \\
\hline
$|\delta Re (C_S)|$ & 7.05 & 1.40 & - & - & 0.20 & - & - & - \\
\hline
$|\delta Re (C_T)|$ & 5.07 & - & 1.00 & - & - & 1.00 & - & - \\
\hline
\end{tabular}}
%\label{dastpol}}
\end{center}
\caption{The results obtained from the analysis of the $D^{\ast}$ polarisation asymmetry in the decay $\bdasttau$.}
\label{dastpol}
%\end{center}
\end{table}

\subsection{ Bin-by-bin analysis}
In general, the sensitivity to various NP interactions may also be $q^2$ dependent. 
Hence, we analyse the bin-by-bin $q^2$ distribution of the differential decay rate of $\bdtau$ to look 
for more possibilities, and zoom in to the regions of $q^2$, within which the sensitivity to a specific type of
new interaction is much larger than most other regions. In general the $\delta C_i$s extracted 
from individual bins are very large, therefore in the figs. \ref{figbins}, \ref{figstbins}
we plot $\bar{\delta C_i} = \delta C_i/N_{norm}$, where $N_{norm}$ is some number used to 
normalised $\delta C_i$.

% \subfloat[ ]
\begin{figure*}[!htbp]
\subfloat[Case $b$ with $C_2 = |C_S|^2$, $C_3 = Im(C_T)^2$, $C_4=Re(C_S)$, and $N_{norm} = 10^5$.]
{\includegraphics[scale=0.32]{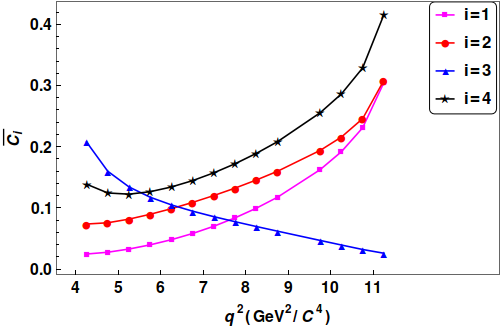}\label{binb}}~~~~
\subfloat[Case $c$ with $C_2 = Im(C_S)^2$, $C_3 = |C_T|^2$, $C_5 =Re(C_T)$, and $N_{norm} = 10^5$.]
{\includegraphics[scale=0.32]{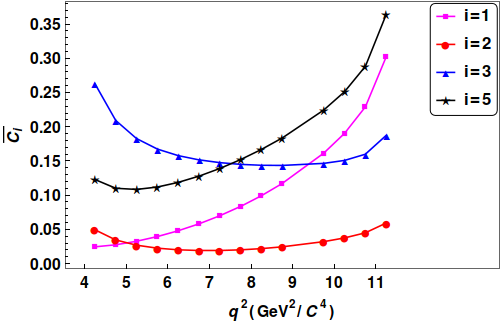}\label{binc}}~~~~
\subfloat[Case $d$ with $C_2 = Im(C_S)^2$, $C_3 = Im(C_T)^2$, and $N_{norm} = 10^4$.]
{\includegraphics[scale=0.32]{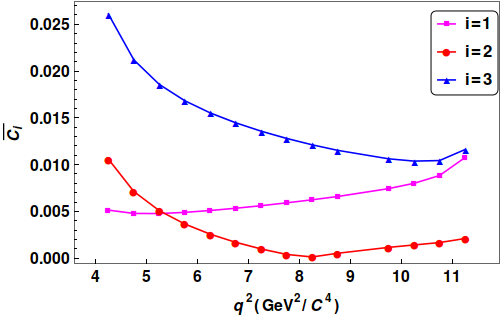}\label{bind}}\\
\subfloat[Case $e$ with $C_2 = |C_S|^2$, $C_4=Re(C_S)$, and $N_{norm} = 10^4$.]
{\includegraphics[scale=0.32]{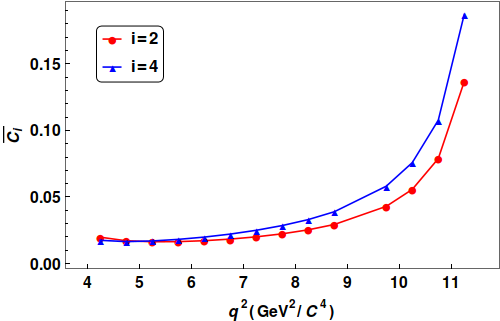}\label{bine}}~~~~
\subfloat[Case $f$  with $C_3 = |C_T|^2$, $C_5=Re(C_T)$, and $N_{norm} = 10^4$]
{\includegraphics[scale=0.32]{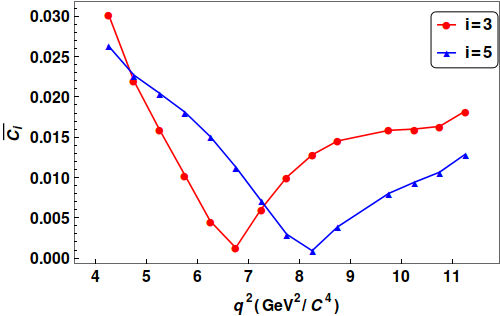}\label{binf}} \\
\caption{Selected cases in $\bd0tau$, here, $\overline{C_i} = |\delta C_i|/N_{norm}$.}
\label{figbins}
\end{figure*} 

The results obtained from the analysis of the $q^2$ distribution of differential decay rate in $\bd0tau$ are presented 
in fig. \ref{figbins}, where the variations of the $\delta C_i$s with  
$q^2$ are shown. The normalised uncertainties in the simultaneous extraction of 
$|C_S|^2$, $|C_T|^2$, $Re(C_T)$ and $Re(C_S)$, and their variations with $q^2$ are shown in 
figs. \ref{binb} and \ref{binc} respectively. On the other hand the variations of $\delta |C_S|^2$, 
and $\delta Re(C_S)$ with $q^2$ when $C_T =0$ are shown in fig. \ref{bine}, while that 
for $\delta |C_T|^2$, and $\delta Re(C_T)$ when $C_S=0$ are shown in fig. \ref{binf}.  
We note that in the low $q^2$ region  ($\lsim 7 {\it GeV}^2/c^4$) the differential decay rate is sensitive 
to the scalar interaction \footnote{In very low $q^2$ regions the $q^2$ distribution is also sensitive 
to $C_1$.}, the sensitivity to tensor interaction in this region is very weak, whereas 
in the high $q^2$ region ($\gsim 7 {\it GeV}^2/c^4$) it is rather sensitive to the tensor interaction. 
In case the Wilson coefficients are purely imaginary, in all the $q^2$ regions, 
the decay rate distribution is sensitive more to the scalar interaction (fig. \ref{bind}) than the others.      

%\begingroup
% \squeezetable

As we noted earlier, $R(D^*)$ is equally sensitive to $C_{V_2}$ and $C_T$ (case $a^*$), however, the analysis 
of the differential decay rate distributions show that (figs. \ref{binkst} and \ref{binist}) it is sensitive 
to tensor interaction only in the very high and low $q^2$ regions, and it is sensitive to $C_{V_2}$ in all the 
$q^2$ regions except the very low $q^2$ region. In fig. \ref{bincst} the variations of $\delta Re(C_{V_2})$,  
$\delta Re(C_S)$, and $\delta |C_S|^2$ with the $q^2$ in the case $C_T=0$ are shown, we note that in all 
the $q^2$ regions the decay rate 
is equally sensitive to vector and scalar interactions except in the very low $q^2$ region, where the sensitivity 
to $Re(C_S)$ is better than that to $Re(C_{V_2})$. We also study the cases when the NP interaction is 
scalar type. The $q^2$ distribution of the extracted errors on the respective parameters is 
shown in fig. \ref{binhst}, which indicates that the decay rate is sensitive 
to scalar interactions only in the low $q^2$ region. If $C_S$ and $C_T$ are purely imaginary than the $q^2$ distribution 
of the decay rate is sensitive to the tensor interactions in all the $q^2$ regions (fig. \ref{binjst}).

\begin{figure*}[!htbp]
\subfloat[Case $c^*$ with $C_2 = Re(C_{V_2})$, $C_3 = |C_S|^2$, $C_5=Re(C_S)$, and $N_{norm} = 10^6$.]
{\includegraphics[scale=0.33]{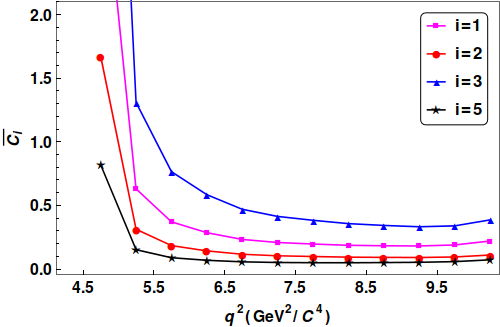}\label{bincst}}~~~
\subfloat[Case $j^*$ with $C_2 = Re(C_{V_2})$, $C_3 = Im(C_T)^2$, $C_4=Im(C_S)^2$, and $N_{norm} = 10^6$.]
{\includegraphics[scale=0.33]{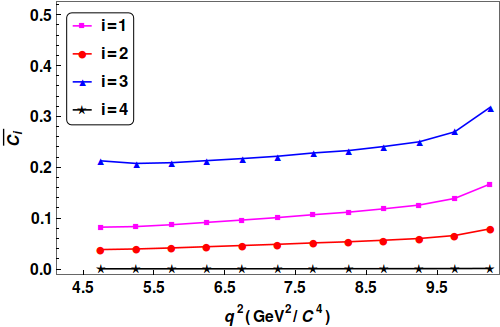}\label{binjst}}~~~
\subfloat[Case $h^*$ with $C_3 = |C_S|^2$, $C_5=Re(C_S)$, and $N_{norm} = 10^4$ ]
{\includegraphics[scale=0.33]{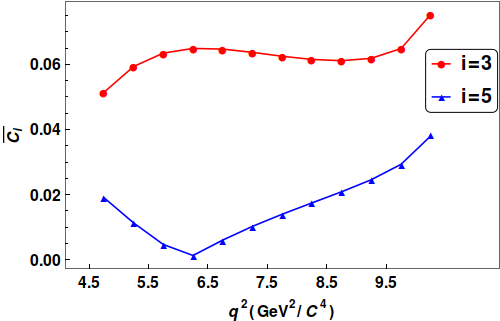}\label{binhst}}\\
\subfloat[Case $i^*$ with $C_4 = |C_T|^2$, $C_6=Re(C_T)$, and $N_{norm} = 10^4$ ]
{\includegraphics[scale=0.33]{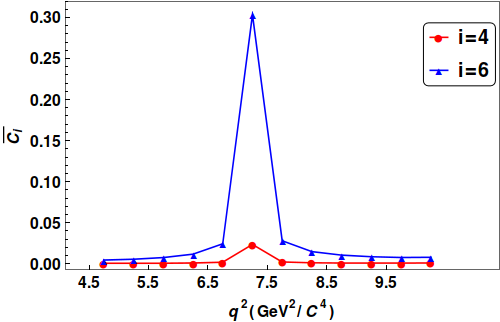}\label{binist}}~~~
\subfloat[Case $k^*$ with $C_2 = Re(C_{V_2})$ and $N_{norm} = 10^4$  ]
{\includegraphics[scale=0.33]{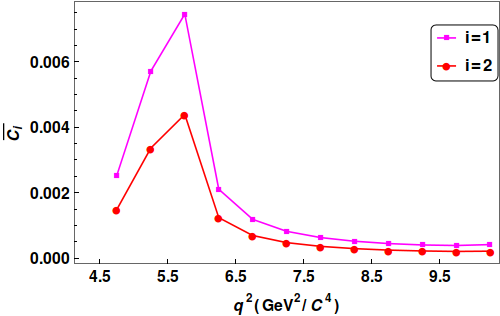}\label{binkst}}~~~
\caption{Selected cases in $\bdasttau$, here, $\overline{C_i} = |\delta C_i|/N_{norm}$.  }
\label{figstbins}
\end{figure*}

All these studies suggest that we could gain in 
NP sensitivity if we focus on specific $q^2$ regions, which we may lose in the full $q^2$-integrated observables.
We note that the sensitivity to a particular type of interaction is limited to particular regions of 
$q^2$. Therefore, the experimental data in the specific regions of $q^2$ could help us a better interpretation of the type
of NP interactions, which may not be obtainable from $q^2$ integrated observables. As for example, if we see large 
deviations in data only in the very high and very low $q^2$ regions, that can be interpreted as due to the presence of 
a tensor interaction (fig. \ref{binist}). On the other hand, if we see deviations only in the low $q^2$ ($< 7$) bins, 
that could be due to a new scalar interaction (fig. \ref{binhst}). Finally, if data shows deviations in most of the 
$q^2$ bin except the very low $q^2$, this can be due to the presence of new vector interaction (fig. \ref{binkst}).  

\begin{figure*}[!htbp]
\subfloat[ ]
{\includegraphics[scale=0.50]{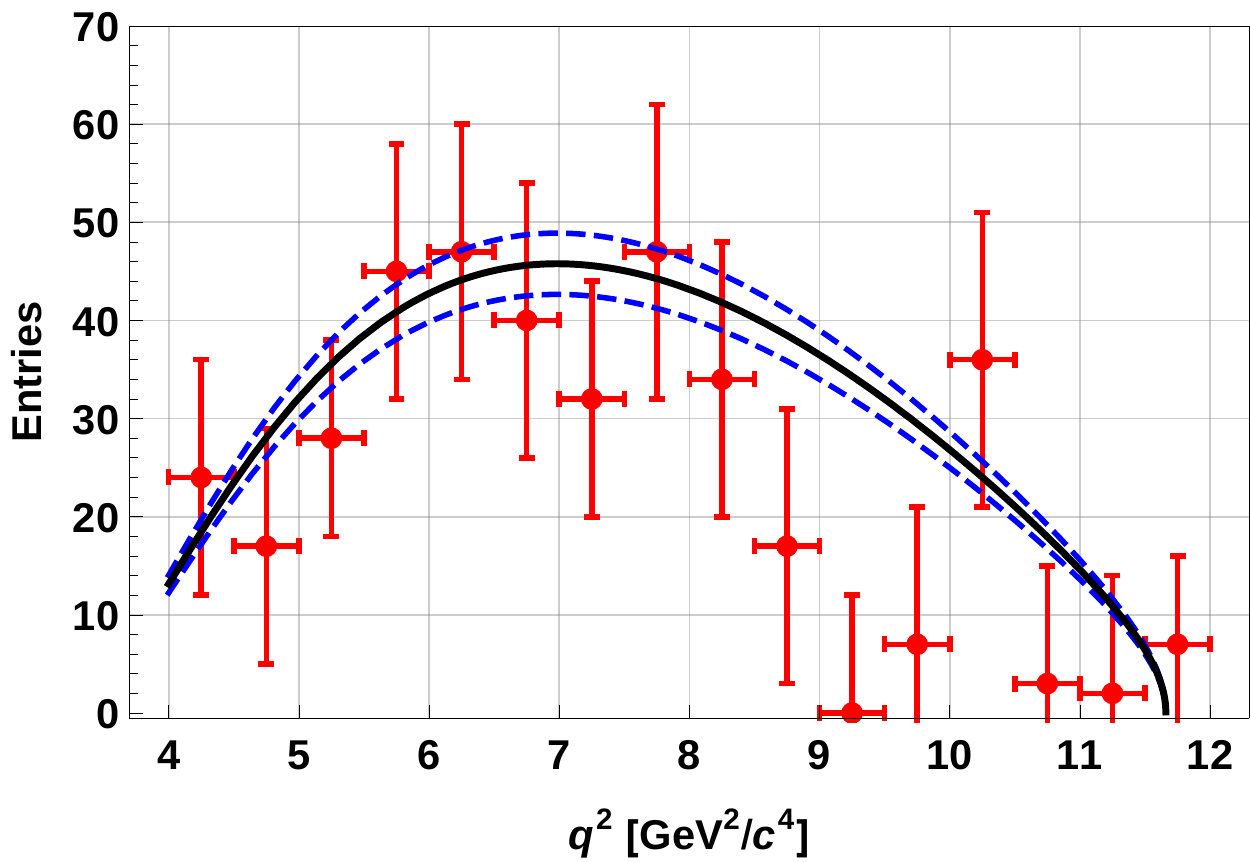}\label{rdbin}}~~~
\subfloat[ ]
{\includegraphics[scale=0.50]{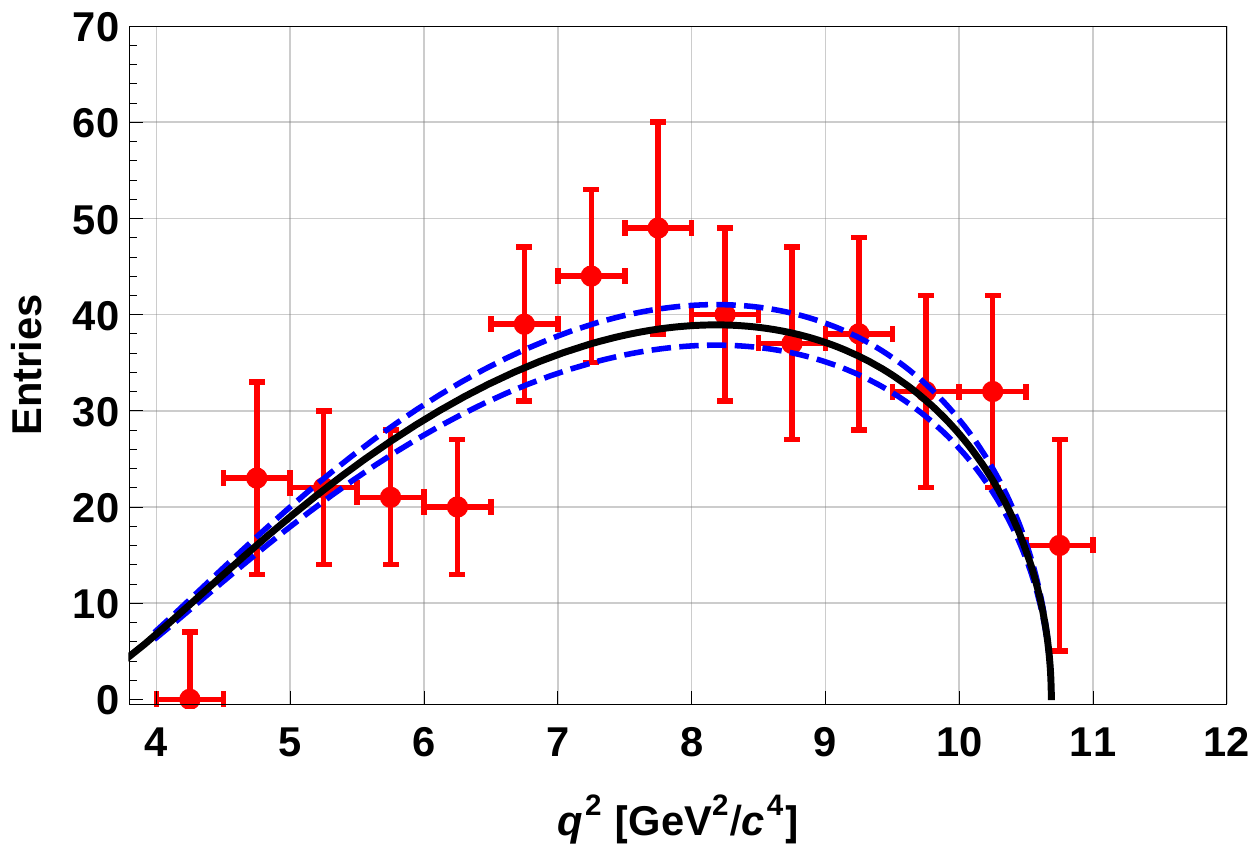}\label{rdastbin}} \\
%\label{binbin}
\caption{Measured $q^2$ distribution of the events in $\bd0tau$ (left) and $\bdasttau$ (right) decays \cite{Lees:2013uzd}.
The black lines represent the $q^2$ distribution predicted in the SM, obtained from the respective
differential branching fractions.}
\label{binbin}
\end{figure*}

In order to explain our point we take the example of the $q^2$ distributions of the measured events in $\bdtau$, which are 
shown in Fig.(\ref{binbin}). The plots are generated using the data given in ref. \cite{Lees:2013uzd,Aubert:2009,Aubert:2007}. 
The predicted $q^2$ distributions in SM with the central values of the form factors are shown by black lines, 
blue dotted lines represent the errors in SM. In Fig.(\ref{rdbin}), we see that the data is not fully consistent 
with the SM in the region $8.0 < q^2 < 11$, which is the region where the decay rate distribution is sensitive to 
tensor interaction, as analysed above in the decay $\bd0tau$. At the moment it is hard to conclude anything, and we 
have to wait for better statistics. From Fig.(\ref{rdastbin}) we see distinct regions of $q^2$ where 
the data is not fully consistent with the SM prediction, and our analysis suggests that those regions could 
potentially be very sensitive to NP effects. Again, because of poor statistics it is premature to conclude anything. 
Therefore, the experimental effort should be in gathering more statistics in
specific regions of $q^2$ potentially sensitive to NP, which may in turn help the clean extraction of NP couplings.

As mentioned above, the numerical estimates are done with central values of all the relevant parameters taken from
\cite{formfactors}. Numerical instability of our results could be main source of uncertainty in our estimates. 
The numerical results of $X_{ij}$, which depend on the matrix inversion of $M_{ij}$, are often unstable; even a tiny
variation of $M_{ij}$ could change $X_{ij}$ significantly. This is why, when we estimate the 
statistical uncertainties in simultaneous extractions of the Wilson coefficients, we allowed only stable solutions. 
We calculate the selected $\delta C_i$ first to $m^{th}$ and then to $(m-1)^{th}$ decimal places, and obtain 
${\delta C_i}^{[m]}$ and ${\delta C_i}^{[(m-1)]}$ respectively. We consider the results as stable only when 
$({\delta C_i}^{[m]} - {\delta C_i}^{[(m-1)]})/{\delta C_i}^{[m]} < 0.01$. We checked the stability up to $m=10$,
and in most of the cases presented above, our results are very much stable, and the error due to 
this is negligible. As we can see from the expressions of $\delta C_i$, the other sources of errors 
in our estimates are given by the errors in $f_i(q^2)$, $\sigma_P$ , ${\cal L}_{eff}$ , and ${\cal B}_{\ell}$. 
It is straight forward to estimate the errors due to $\sigma_P$ , ${\cal L}_{eff}$ , and ${\cal B}_{\ell}$. 

\begin{table}[!htbp]
 \begin{center}
% \subfloat[ ]
{\begin{tabular}{ c c | c | c | c | c |}
%\cline{3-6} & & \multicolumn{4}{ c |}{$|\delta C_i| /|\delta C_1|(\sqrt{\delta C_i})$}\\
%\cline{3-6}
\cline{1-6}
\multicolumn{1}{|c}{Cases}& \multicolumn{1}{|l|}{}
&$\delta C_i$  & ${\delta C_i}^+$ &  ${\delta C_i}^-$ & $\pm \% Err.  $ \\
\cline{1-6}\multicolumn{1}{|c}{\multirow{3}{*}{d} } & \multicolumn{1}{| c |}{$C_1$} & 0.082 & 0.079 & 0.087 & 5.032 \\
\cline{2-6}\multicolumn{1}{| c }{}& \multicolumn{1}{| c |}{$Im(C_S)^2$ } & 0.013 & 0.013 & 0.014 & 5.032 \\
\cline{2-6}\multicolumn{1}{| c }{}& \multicolumn{1}{| c |}{$Im(C_T)^2$ } & 0.168 & 0.160 & 0.177 & 5.032 \\
\hline\multicolumn{1}{|c}{\multirow{3}{*}{e} } &  \multicolumn{1}{| c |}{$C_1$} & 0.240 & 0.229 & 0.253 & 5.031 \\
\cline{2-6}\multicolumn{1}{| c }{}& \multicolumn{1}{| c |}{$|C_S|^2$}& 0.274 & 0.261 & 0.289 & 5.031 \\
\cline{2-6}\multicolumn{1}{| c }{}& \multicolumn{1}{| c |}{$Re(C_S)$}& 0.354 & 0.337 & 0.373 & 5.031 \\
% \hline\multicolumn{1}{|c}{\multirow{2}{*}{d} } &  \multicolumn{1}{| c |}{9.5 - 11.5} & 0.94 & - & 1.28 & - \\
% \cline{2-6}\multicolumn{1}{| c }{}& \multicolumn{1}{| c |}{full region }& 1.14 & - & 1.47 & - \\
% \hline\multicolumn{1}{|c}{\multirow{3}{*}{e} } & \multicolumn{1}{| c |}{6.0 - 7.0 } & - & 0.2 & - & 1.21 \\
% \cline{2-6}\multicolumn{1}{| c }{}& \multicolumn{1}{| c |}{7.5 - 9.0 } & - & 2.29 & - & 0.29(1.07) \\
% \cline{2-6}\multicolumn{1}{| c }{}& \multicolumn{1}{| c |}{full region } & - & 3.27 & - & 1.66 \\
\hline
\end{tabular}}
%\caption{ Numerical values of $\delta C_i$s, considering $\pm 10\%$ errors in $f_i(q^2)$ for different cases of $R(D)$}
\end{center}
\caption{ Numerical values of $\delta C_i$s, and ${\delta C_i}^+$ (${\delta C_i}^-$)
considering +10\% (-10\%) errors in $f_i(q^2)$ for few cases of $R(D)$. The \% error is given by 
$(\delta C_i - {\delta C_i}^{\pm})/{\delta C_i}$.}
\label{err1}
\end{table}
However, the estimate due to $f_i(q^2)$s are not that straight forward since $M_{ij}$s depend solely 
upon them. The main sources of uncertainties in $f_i$s, including the SM, are the form-factors. 
Errors due to other parameters, like CKM element etc, are canceled in the ratios. 
In the tables \ref{err1}, and \ref{err2}, we consider a few cases and give a rough estimate of the 
uncertainties due to the errors in $f_i(q^2)$. The overall error is about $\pm$ 5\% in the extraction of $\delta C_i$s,
if we consider the errors in $f_i(q^2)$s are about $\pm$ 10\%. We also estimate the errors in $\delta C_i$ 
by considering the actual errors in all the form-factors given in ref. \cite{formfactors}, and find that they are even 
smaller than whatever we have shown in the above mentioned tables. Finally, we would like to comment that the estimated errors 
due to form-factors, and the other experimental parameters will have almost equal impact on all the extracted $\delta C_i$s,
which is also small. Therefore, our conclusions about the relative sensitivities will not change.

\begin{table}[!htbp]
 \begin{center}
% \subfloat[ ]
{\begin{tabular}{ c c | c | c | c | c |}
%\cline{3-6} & & \multicolumn{4}{ c |}{$|\delta C_i| /|\delta C_1|(\sqrt{\delta C_i})$}\\
%\cline{3-6}
\cline{1-6}
\multicolumn{1}{|c}{Cases}& \multicolumn{1}{|l|}{}
&$\delta C_i$  & ${\delta C_i}^+$ &  ${\delta C_i}^-$ & $\pm \% Err. $ \\
\cline{1-6}\multicolumn{1}{|c}{\multirow{3}{*}{$f^*$} } & \multicolumn{1}{| c |}{$C_1$} & 0.448 & 0.427 & 0.472 & 5.032 \\
\cline{2-6}\multicolumn{1}{| c }{}& \multicolumn{1}{| c |}{$Re(C_{V_2})$ } & 0.211 & 0.201 & 0.222 & 5.032 \\
\cline{2-6}\multicolumn{1}{| c }{}& \multicolumn{1}{| c |}{$Im(C_T)^2$ } & 0.006 & 0.0058 & 0.0065 & 5.032 \\
\hline\multicolumn{1}{|c}{\multirow{3}{*}{$g^*$} } &  \multicolumn{1}{| c |}{$C_1$} & 0.042 & 0.040 & 0.044 & 5.032 \\
\cline{2-6}\multicolumn{1}{| c }{}& \multicolumn{1}{| c |}{$Im(C_S)^2$}& 0.961 & 0.916 & 1.013 & 5.032 \\
\cline{2-6}\multicolumn{1}{| c }{}& \multicolumn{1}{| c |}{$Im(C_T)^2$}& 0.0048 & 0.0046 & 0.0051 & 5.032 \\
% \hline\multicolumn{1}{|c}{\multirow{2}{*}{d} } &  \multicolumn{1}{| c |}{9.5 - 11.5} & 0.94 & - & 1.28 & - \\
% \cline{2-6}\multicolumn{1}{| c }{}& \multicolumn{1}{| c |}{full region }& 1.14 & - & 1.47 & - \\
% \hline\multicolumn{1}{|c}{\multirow{3}{*}{e} } & \multicolumn{1}{| c |}{6.0 - 7.0 } & - & 0.2 & - & 1.21 \\
% \cline{2-6}\multicolumn{1}{| c }{}& \multicolumn{1}{| c |}{7.5 - 9.0 } & - & 2.29 & - & 0.29(1.07) \\
% \cline{2-6}\multicolumn{1}{| c }{}& \multicolumn{1}{| c |}{full region } & - & 3.27 & - & 1.66 \\
\hline
\end{tabular}}
%\caption{ Numerical values of $\delta C_i$s, considering $\pm 10\%$ errors in $f_i(q^2)$ for different cases of $R(D^*)$}
\end{center}
\caption{ Numerical values of $\delta C_i$s, and ${\delta C_i}^+$ (${\delta C_i}^-$)
considering +10\% (-10\%) errors in $f_i(q^2)$ for few cases of $R(D^*)$. The \% error is given by 
$(\delta C_i - {\delta C_i}^{\pm})/{\delta C_i}$.}
\label{err2}
\end{table}

\section{Summary}
We use the optimal observable technique to test the sensitivities of the various observables in $\bdtau$ 
to the various NP interactions, like new vector, scalar and tensor interactions. Numerically, we find that 
the observables in $\bd0tau$ are more or less equally sensitive to scalar and tensor interactions, only exception 
is the $\tau$ polarisation asymmetry, where $\delta Re(C_S) < \delta Re(C_T)$ but they are of same order. 
Therefore, even if the measured values of the observables deviate from their SM expectations, a priori it would be 
hard to decide what type of new interaction will it be. However, the analysis of the $q^2$ distribution of 
the decay rate allows us to separate the regions of $q^2$ which are sensitive to scalar interaction (low $q^2$) and 
tensor interaction (high $q^2$). 

The overall sensitivity of the observables in $\bdasttau$ is more towards tensor interactions, in particular to $|C_T|^2$. 
Also, we note that $\delta Re(C_T) < \delta Re(C_{V_2})$ but they are of same order, hence, 
we need better statistics to distinguish their effects. The decay $\bdasttau$ has very poor sensitivity to 
scalar interaction compared to the tensor interaction, the only exceptions being the $D^*$ polarisation, $A_{FB}^{R^*}$. 
However, in the absence of tensor interactions, the decay $\bdasttau$ is equally sensitive to real part of 
both the vector and scalar Wilson coefficients, sensitivity to $|C_S|^2$ is much less 
compared to the real parts. However, the analysis of the $q^2$ distributions of the decay rate shows 
distinct regions of $q^2$, which are sensitive to vector, scalar, and tensor interactions respectively. 
These sensitivities are lost in the full $q^2$ integrated observables. Present data on different bins 
do not have sufficient statistics to conclude anything, more precise data could help us to pinpoint 
the type of NP interaction. Therefore, in an experiment, the priority should be given to gaining
statistics at those regions of $q^2$.

We note that both the decay modes are more sensitive to the real part of the coefficients compared to 
imaginary part. Among the various observables, $\tau$ polarisation asymmetries 
have better sensitivity to the relevant new coefficients ($C_i$); the uncertainties on the extracted $C_i$s are either 
less or comparable to that obtained in others. Therefore, future data on $\tau$ polarisation asymmetries could put
tighter constraints on the NP parameter space.

\section{Acknowledgement}
We would like to thank Subhaditya Bhattacharya for useful discussions on optimal-observable analysis. SN would also like to
thank Thomas Kuhr and Bipul Bhuyan for useful discussions. 

\begin{widetext}
\section{Appendix}

In the following tables, the various $f_i(q^2)$ for different observables used in our analysis are shown.

\begin{table*}[!htbp]
\begin{center}
{\begin{tabular}{| c | c | c |}
\hline
\backslashbox{$f_i$}{Obs} & $R(D)$ & $P_\tau ^R$ \\
\hline
$f_1$ & $\mathcal{G}\left(\left(1+\frac{m_\tau^2}{2q^2}\right){H^{s}_{V,0}}^2+\frac{3}{2}\frac{m_\tau^2}{q^2}{H^{s}_{V,t}}^2\right)$ &$ \mathcal{G}\left(\left(-1+\frac{m_\tau^2}{2q^2}\right){H^{s}_{V,0}}^2+\frac{3}{2}\frac{m_\tau^2}{q^2}{H^{s}_{V,t}}^2\right)$\\
\hline
$f_2$ & $\frac{3}{2}\mathcal{G}{H^{s}_S}^2$ & $\frac{3}{2}\mathcal{G}{H^{s}_S}^2$  \\
\hline
$f_3$ & $8\mathcal{G}\left(1+\frac{2m_\tau^2}{q^2}\right){H^{s}_T}^2$ & $ 8\mathcal{G}\left(1-\frac{2m_\tau^2}{q^2}\right){H^{s}_T}^2$ \\
\hline
$f_4$ & $3\mathcal{G}\frac{m_\tau}{\sqrt{q^2}}H^s_S H^s_{V,t}$ & $3\mathcal{G}\frac{m_\tau}{\sqrt{q^2}}H^s_S H^s_{V,t}$ \\
\hline
$f_5$ & $-12 \mathcal{G}\frac{m_\tau}{\sqrt{q^2}}H^s_T H^s_{V,0}$ & $  4 \mathcal{G}\frac{m_\tau}{\sqrt{q^2}}H^s_T H^s_{V,0} $ \\
\hline
\end{tabular}}\\
\end{center}
%\label{fisrd}
\caption{$f_i$s for $R_D$ and $\tau$ polarisation asymmetry in $\bd0tau$.}
\label{fisrd}
%\end{center}
\end{table*}

\begin{table*}[!htbp]
\begin{center}
{\begin{tabular}{| c | c | c |}
\hline
\backslashbox{$f_i$}{Obs} & $R(D^*)$ & $P_\tau ^{R^*}$ \\
\hline
$f_1$ & $\mathcal{G^*}\left( \left(1+\frac{m_\tau^2}{2q^2}\right)\left(H^2_{V,+}+H^2_{V,-}+H^2_{V,0}\right)+\frac{3}{2}\frac{m_\tau^2}{q^2}H^2_{V,t}\right)$ & $\mathcal{G^*}\left( \left(-1+\frac{m_\tau^2}{2q^2}\right)\left(H^2_{V,+}+H^2_{V,-}+H^2_{V,0}\right)+\frac{3}{2}\frac{m_\tau^2}{q^2}H^2_{V,t}\right)$\\
\hline
$f_2$ & $-2\mathcal{G^*}\left( \left(1+\frac{m_\tau^2}{2q^2}\right)\left(H^2_{V,0}+2H_{V,+}H_{V,-}\right)+\frac{3}{2}\frac{m_\tau^2}{q^2}H^2_{V,t}\right)$ & $\mathcal{G^*}\left( \left(2-\frac{m_\tau^2}{q^2}\right)\left(H^2_{V,0}+2H_{V,+}H_{V,-}\right)-3\frac{m_\tau^2}{q^2}H^2_{V,t}\right)$ \\
\hline
$f_3$ & $\frac{3}{2}\mathcal{G^*}H^2_S$& $\frac{3}{2}\mathcal{G^*}H^2_S$\\
\hline
$f_4$ & $8\mathcal{G^*}\left(\left(1+\frac{2m_\tau^2}{q^2}\right)\left(H^2_{T,+}+H^2_{T,-}+H^2_{T,0}\right)\right)$ & $8\mathcal{G^*}\left( \left(1-\frac{2m_\tau^2}{q^2}\right)\left(H^2_{T,+}+H^2_{T,-}+H^2_{T,0}\right)\right)$\\
\hline
$f_5$ & $3\mathcal{G^*}\frac{m_\tau}{\sqrt{q^2}}H_SH_{V,t}$ & $3\mathcal{G^*}\frac{m_\tau}{\sqrt{q^2}}H_SH_{V,t}$\\
\hline
$f_6$ & $-12\mathcal{G^*} \frac{m_\tau}{\sqrt{q^2}}\left(H_{T,0}H_{V,0}+H_{T,+}H_{V,+}-H_{T,-}H_{V,-}\right)$ & $4\mathcal{G^*} \frac{m_\tau}{\sqrt{q^2}}\left(H_{T,0}H_{V,0}+H_{T,+}H_{V,+}-H_{T,-}H_{V,-}\right)$\\
\hline
$f_7$ & $12 \mathcal{G^*}\frac{m_\tau}{\sqrt{q^2}}\left(H_{T,0}H_{V,0}+H_{T,+}H_{V,-}-H_{T,-}H_{V,+}\right)$ & $-4 \mathcal{G^*}\frac{m_\tau}{\sqrt{q^2}}\left(H_{T,0}H_{V,0}+H_{T,+}H_{V,-}-H_{T,-}H_{V,+}\right)$ \\
\hline
\end{tabular}}\\
\end{center}
\label{fisrdast}
\caption{$f_i$s for $R_{D^*}$ and $\tau$ polarisation asymmetry in $\bdasttau$.}
\label{fisrdast}
%\end{center}
\end{table*}
\begin{table*}[!htbp]
\begin{center}
\begin{tabular}{|| c | c | c | c ||}
\hline
\backslashbox{$f_i$}{Obs} & $\mathcal{A^R_{FB}}$ & $\mathcal{A^{R^*}_{FB}}$ & $P^R_{D^*}$\\
\hline
$f_1$ & $\mathcal{F}\left(\frac{m_\tau^2}{q^2}H^s_{V,0}H^s_{V,t}\right)$ & $\frac{1}{2}\mathcal{F^*}\left(H^2_{V,+}-H^2_{V,-}\right)$ & $\mathcal{G^*}\left(\left( 1 + {m_\tau^2 \over2q^2} \right) H_{V,0}^2 + {3 \over 2}{m_\tau^2 \over q^2} \, H_{V,t}^2 \right)$\\
\hline
$f_2$ & $\mathcal{F}\left(\frac{m_\tau}{\sqrt{q^2}}H^s_{V,0}H^s_S\right)$ & $\mathcal{F^*}\frac{m_\tau^2}{q^2}H_{V,0}H_{V,t}$ & ${3 \over 2}\mathcal{G^*}H_S^2$\\
\hline
$f_3$ & $-4\mathcal{F}\left(\frac{m_\tau}{\sqrt{q^2}}H^s_{V,t}H^s_T\right)$ & $8\mathcal{F^*}\frac{m_\tau^2}{q^2}\left(H^2_{T,+}-H^2_{T,-}\right)$ & $8\mathcal{G^*}\left( 1+ {2m_\tau^2 \over q^2} \right) H_{T,0}^2$\\
\hline
$f_4$ & $-4\mathcal{F}H^s_SH^s_T $ & $\mathcal{F^*}{m_\tau \over \sqrt{q^2}} \, H_S H_{V,0}$ & $3\mathcal{G^*} {m_\tau \over \sqrt{q^2}} \, H_S H_{V,t}$\\
\hline
$f_5$ & $-$ & $-4\mathcal{F^*}{m_\tau \over \sqrt{q^2}} \, \left( H_{T,0} H_{V,t} + H_{T,+} H_{V,+} + H_{T,-} H_{V,-} \right)$ & $-12\mathcal{G^*}{m_\tau \over \sqrt{q^2}} H_{T,0} H_{V,0}$\\
\hline
$f_6$ & $-$ & $4\mathcal{F^*} \frac{m_\tau}{\sqrt{q^2}}\left(H_{T,0}H_{V,t}+H_{T,+}H_{V,-}+H_{T,-}H_{V,+}\right)$ & $-$\\
\hline
$f_7$ & $-$  & $-4 \mathcal{F^*} H_{T,0} H_S$ & $-$\\
\hline
\end{tabular}\\
\end{center}
\caption{$f_i$s for $\tau$ forward-backward asymmetries in $\bdtau$ decays, and $D^{\ast}$ polarisation asymmetry in
$\bdasttau$. }
\label{fisasy}
%\end{center}
\end{table*}
The expressions for ${\cal F}$, ${\cal F}^{\ast}$, ${\cal G}$, ${\cal G}^{\ast}$ are given by \cite{Sakaki:2013}
\bea
\mathcal{G} &=&\frac{\tau_B}{\mathcal{B} (B\to Dl\nu)} \frac{G^2_F{|V_{cb}|}^2}{192\pi^3m_B^3} q^2
\sqrt{\lambda_D(q^2)}{\left(1-\frac{m_\tau^2}{q^2}\right)}^2, \nn \\
\mathcal{G^*}&=&\frac{\tau_B}{\mathcal{B}(B\to D^*l\nu)}\frac{G^2_F{|V_{cb}|}^2}{192\pi^3m_B^3}q^2
\sqrt{\lambda_{D^*}(q^2)}{\left(1-\frac{m_\tau^2}{q^2}\right)}^2 \nn \\
\mathcal{F} &=&\frac{\tau_B}{\mathcal{B}(B\to Dl\nu)}\frac{G^2_F{|V_{cb}|}^2}
{128\pi^3m_B^3}q^2\sqrt{\lambda_{D}(q^2)}{\left(1-\frac{m_\tau^2}{q^2}\right)}^2 \nn \\
\mathcal{F^*} &=&\frac{\tau_B}{\mathcal{B}(B\to D^*l\nu)}\frac{G^2_F{|V_{cb}|}^2}{128\pi^3m_B^3}q^2
\sqrt{\lambda_{D^*}(q^2)}{\left(1-\frac{m_\tau^2}{q^2}\right)}^2
\eea

\end{widetext}

\end{document}